\documentclass[10pt]{iopart}

\usepackage{graphicx}
\usepackage{gensymb}

\begin{document}

\title[Dispersing artifacts in FT-STS]{Dispersing artifacts in FT-STS: a comparison of set point effects across acquisition modes}
\author{A J Macdonald$^{1,2}$, Y-S Tremblay-Johnston$^{2,3}$, S Grothe$^{1,2}$, S Chi$^{1,2}$, P Dosanjh$^{1,2}$, S Johnston$^4$ and S A Burke$^{1,2,3}$}
\address{$^1$Department of Physics and Astronomy, University of British Columbia, Vancouver, BC, Canada V6T 1Z1}
\address{$^2$Stewart Blusson Quantum Matter Institute, University of British Columbia, Vancouver, BC, Canada, V6T 1Z4}
\address{$^3$Department of Chemistry, University of British Columbia, Vancouver, BC, Canada V6T 1Z1}
\address{$^4$Department of Physics and Astronomy, University of Tennessee, Knoxville, Tennessee 37996-1200, USA} 
\ead{saburke@phas.ubc.ca}

\vspace{10pt}
\begin{indented}
\item[]July 2016
\end{indented}

\begin{abstract}
Fourier transform scanning tunnelling spectroscopy (FT-STS), or quasiparticle interference (QPI), has become an influential tool for the study of a wide range of important materials in condensed matter physics.  However, FT-STS in complex materials is often challenging to interpret, requiring significant theoretical input in many cases, making it crucial to understand potential artifacts of the measurement.  Here, we compare the most common modes of acquiring FT-STS data and show through both experiment and simulations that artifact features can arise that depend on how the tip height is stabilized throughout the course of the measurement. The most dramatic effect occurs when a series of dI/dV maps at different energies are acquired with simultaneous constant current feedback; here a feature that disperses in energy appears that is not observed in other measurement modes.  Such artifact features are similar to those arising from real physical processes in the sample and are susceptible to misinterpretation. 
\end{abstract}


\maketitle

\ioptwocol
\pagenumbering{arabic}

\section{Introduction}

Scanning tunnelling microscopy (STM) is a powerful tool for relating structural features to real-space electronic structure on the atomic scale.  However, for materials with delocalized electronic states, the momentum dependent band structure provides an important representation for understanding bulk and surface electronic properties.  This reciprocal space picture can be probed by STM via scattering of the electrons in the material by defects (e.g. impurities, step edges, and adatoms), which establish an interference pattern with scattering wavevectors that connect pieces of the underlying band structure of the material.  By visualizing the Fourier transform of these interference patterns, information about the momentum dependence of the electronic structure becomes accessible, including the ability to map the electronic dispersion of both occupied and unoccupied bands, and locally correlate this with surface structure.  This technique, known as Fourier-transform scanning tunnelling spectroscopy (FT-STS), or quasiparticle interference (QPI), has emerged as an important probe of a wide range of complex materials including superconductors \cite{McElroy:2003ug,He:2014wf,Huang:2015dg,Fujita:2014kg,Chi2014,Chuang:2010tl}, topological insulators \cite{Okada:2011gv,Zeljkovic:2014ef,Beidenkopf:2011jw,Roushan:2016kk,Reis:2016dg,Kohsaka:2015fp,Zhang:2013is}, and graphene \cite{Leicht:2014jh,Rutter:2007ep,Mallet:2012ib}.  With the remarkable stability of today's instrumentation, large data sets can be acquired yielding resolution in energy and momentum space rivaling that of state-of-the-art angle resolved photoemission \cite{Grothe2013} allowing new insight into physical processes such as electron-boson coupling \cite{Grothe2013,Allan:2014iy}.  These advances in FT-STS, combined with the real-space resolution of STM, provide a unique view into the connection between physical and electronic structure, for example where there are different surface terminations with different electronic structure \cite{Kohsaka:2015fp,Allan:2012kh,Inoue1184}.

However, there is a catch; typically one must have some \textit{a priori} knowledge of the underlying band structure to assign meaning to the features observed in FT-STS measurements because this relies on scattering processes from one part of the band structure to another.  Simple 1-band systems can be analyzed directly, and this was first done for the surface states of noble metals \cite{Sprunger1997,Hofmann1997,Petersen1998,Petersen1998_2,Schouteden2009} which have more recently been used to advance FT-STS as a probe for many-body effects \cite{Grothe2013,Sessi2015}.  However, more complicated systems require complementary information about the band structure, and the patterns observed can become extremely complex when scattering between multiple bands \cite{Huang:2015dg,Allan:2012kh,Steinbrecher:2013ff}, and the possibility of selection rules for the allowed scattering processes \cite{Chi2014,Okada:2011gv,Mallet:2012ib,Vishik2009}, come into play.  These numerous features that may or may not disperse in energy can prove difficult to disentangle, usually requiring extensive theoretical support.  In these cases it is all the more important that the data itself is well understood.  Several different methods of acquiring FT-STS data have emerged, which can also influence the observed features. Therefore a careful comparison and analysis of these modes of acquisition and their potential artifacts is essential.  Here, we return to a simple, well-known system: the Ag(111) surface state.  We show through both experiment and theory that the way in which the tip height is maintained throughout the measurement -- ``set point effects'' -- can generate potentially misleading artifacts.

\section{Methods}

The Ag(111) surface supports a well characterized two-dimensional Shockley surface state \cite{Paniago1995,Paniago1995_2,Li1997,Li1998,Burgi2000,Burgi2000_2,Kliewer2000,Eiguren2002,Grothe2013}. We chose this noble metal surface state because it serves as a well-studied, theoretically understood material with a single band, that is ideal for comparison of different FT-STS acquisition modes. Within the energy range we probed, the band dispersion can be well described by a free electron model of the form

\begin{equation}
\epsilon(\textbf{k}) = \frac{\hbar^2 \textbf{k}^2_{\|}}{2m^*} - \mu,
\end{equation}

\noindent where $\epsilon(\bf{k})$ is the band dispersion, $\hbar$ is the reduced Planck constant, $m^*$ is the effective electron mass, which for the Ag(111) surface state is approximately equal to 0.4 times the free electron mass \cite{Grothe2013,Paniago1995_2,Li1997,Burgi2000_2,Nicolay2001,Reinert2001}, and $\epsilon(0) = -\mu = -65$ meV is the chemical potential. 

All measurements were made at 4.2 K in ultrahigh vacuum with a pressure $< 1 \times 10^{-10}$ mbar using a commercial Createc STM. An electrochemically etched tungsten tip was used for all measurements, and was further prepared \emph{in situ} by sputtering and annealing to remove the oxide layer.  Initial contact with the Ag crystal likely results in an Ag terminated tip. The Ag(111) crystal was cleaned by 3 cycles of sputtering under $2.0 \times 10^{-5}$ mbar Ar atmosphere and annealing at $500 \degree$C to produce large, clean terraces with a low density of CO adsorbates that act as scattering centers (see Figure 1a). Spatial calibration was performed once prior to all measurements by obtaining atomic resolution of the Ag(111) surface to ensure accurate real and reciprocal space measurements.  Two main methods of collecting spatially resolved spectroscopic data were used.  I-V spectroscopic ÒgridsÓ were measured with varying size and spatial resolution (giving different pixel densities) and a thermally limited energy resolution of 1.5 meV. Each I-V spectrum consisted of 512 data points, which were Gaussian smoothed over 3 adjacent points in energy and averaged over 8-12 repeated measurements at each spatial location. The dI/dV was calculated numerically from these processed I-V spectra, giving dI/dV(x,y,V). Typical grid measurements took between 50 and 80 hours to complete. The dI/dV ÒmapsÓ were taken using a lock-in amplifier with a bias modulation frequency of 1.017 kHz and amplitude of 5 mV.  As these yield only one energy, a series of maps was acquired to investigate dispersion.  The dI/dV map measurements were acquired with a simultaneous constant current topography, except for constant height measurements where the feedback was disengaged.  Throughout the rest of the paper we will refer to these methods as ``grids'',``maps'', or ``constant height maps'' if no feedback was used.

Once spectroscopic data is acquired, the FT-STS is obtained from the dI/dV image for each energy.  Figure 1b) shows the real-space dI/dV image extracted from a grid measurement at the Fermi energy ($E_F$) for the region corresponding to the topography shown in Figure 1a).  Surface state scattering from CO adsorbates and step edges is clearly visible.  Since the CO adsorbates and step edges exhibit different scattering potentials, in order to attain a clean FT-STS, a real space correction is applied replacing the step edges (region between dashed lines in Figure 1b) with the average intensity of the dI/dV image. Figure 1c) then shows the Fourier transform of this dI/dV map into scattering or \textbf{q}-space. The bright ring is indicative of intraband scattering across the surface state band of Ag(111) and has a radius \textbf{q}$_F$ = 2\textbf{k}$_F$ at $E_F$. Since the intraband surface state scattering is isotropic we can perform an angular average to improve the signal to noise of the scattering intensity, S(q$_r$), where q$_r$ denotes the radial component of \textbf{q}.  Additional intensity at the top and bottom of this ring is due to the non-isotropic scattering off of the step edges remaining after the real space filtering. To further reduce the influence of this different scatterer we take a restricted angular average only over the left and right quadrants between the dashed lines to produce Figure 1d): the azimuthally averaged line cut of the Fourier transformed data. The peak in Figure 1d) corresponds to the Fermi wave vector q$_F = 0.168 \pm 0.002$ \AA$^{-1}$. By performing the process highlighted in Figure 1a)-d) at multiple energies it is possible to measure the scattering intensity as a function of energy and q$_r$, S(q$_r$,E), recovering the dispersion $\epsilon(q_r)$ (see Figure 2). These steps were also used in Grothe \emph{et al.} \cite{Grothe2013}, and similar processing was performed on all of the data presented here as needed.

\begin{figure}[htbp]
\begin{center}
\includegraphics[width=8.5cm]{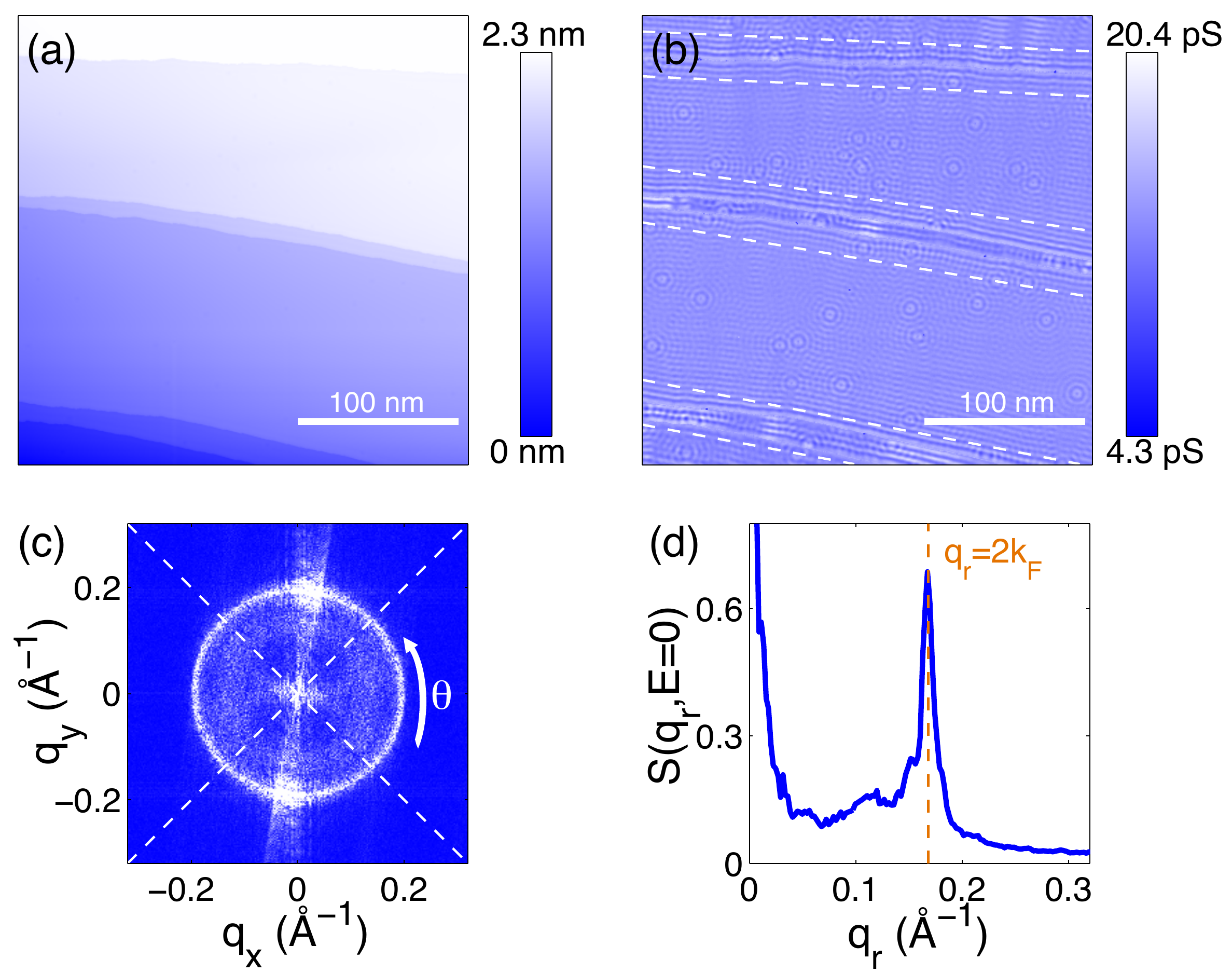}
\caption{Typical FT-STS data showing how the analysis proceeds from real space to \textbf{q}-space from a spectroscopic grid ($V_s = -40$ meV, $I_t=540 $ pA, 282 nm x 282 nm). (a) Real space topography showing multiple terraces separated by step edges. (b) Conductance map (dI/dV) at $E_F$ (V = 0); the white dashed lines indicate where step edges have been removed before taking the Fourier transform. (c) Absolute value of the Fourier transform of the dI/dV showing a dominant ring resulting from intra-band scattering of the Ag(111) surface state. Here the dashed lines show the restricted angular average in \textbf{q}-space. (d) The angular average of the absolute Fourier transform signal. The primary feature at \textbf{q}$_F$ = 2\textbf{k}$_F$ corresponds to the scattering vector of the surface state at $E_F$. The smaller feature at lower q$_r$ is a set point driven effect.}
\label{Fig1}
\end{center}
\end{figure}

The two common methods described above for obtaining the dI/dV(x,y,V) needed to construct the FT-STS scattering dispersions differ in how the tip height is maintained. For grid measurements, where a full STS is acquired at each pixel, the tip height is usually stabilized at each point using consistent tunnelling parameters throughout to account for drift over the long timescale of the measurement. For dI/dV maps acquired with a lock-in amplifier at multiple energies, a constant current feedback is usually used to maintain the tip-sample distance and generate a simultaneous topographic image with each energy mapped.  

\section{Results}

Figure 2 shows the scattering intensity S(q$_r$,E) measured by FT-STS in each of the two main acquisition modes. Both measurements show the expected parabolic dispersion of the surface state band, but differ in other features.  As determined previously in Grothe \emph{et al}. [17], the intensity below the onset of the band is a product of the nature of the scatterer and varies somewhat in intensity between measurements regardless of acquisition mode depending on whether the dominant scatter is CO, other impurities, or if there is significant intensity from step edge scattering that remains after filtering. We therefore do not focus on this feature here. Instead we examine the dispersive QPI features related to the dispersing band structure of the system. Most notably, the grid measurement (Figure 2a) shows a faint, broad vertical feature slightly above \textbf{q}$_F$ = 2\textbf{k}$_F$ in addition to the expected intraband scattering. Whereas for the measurement made by acquiring dI/dV maps at different energies, an additional faint, and also somewhat broad feature appears that instead disperses, crossing \textbf{q}$_F$ = 2\textbf{k}$_F$ at $E_F$.  The dI/dV maps (Figure 2b) also show a strongly varying background intensity as a function of energy, strongest near $E_F$, that is not seen in the grid measurement. The constant current maps have lower energy resolution compared to grid measurements over the same time scale, due to a difference in the speed of data acquisition between the two measurement techniques.  We can now clearly see that acquiring the FT-STS dispersion using different methods yields qualitatively different results.

\begin{figure}[htbp]
\begin{center}
\includegraphics[width=8.4cm]{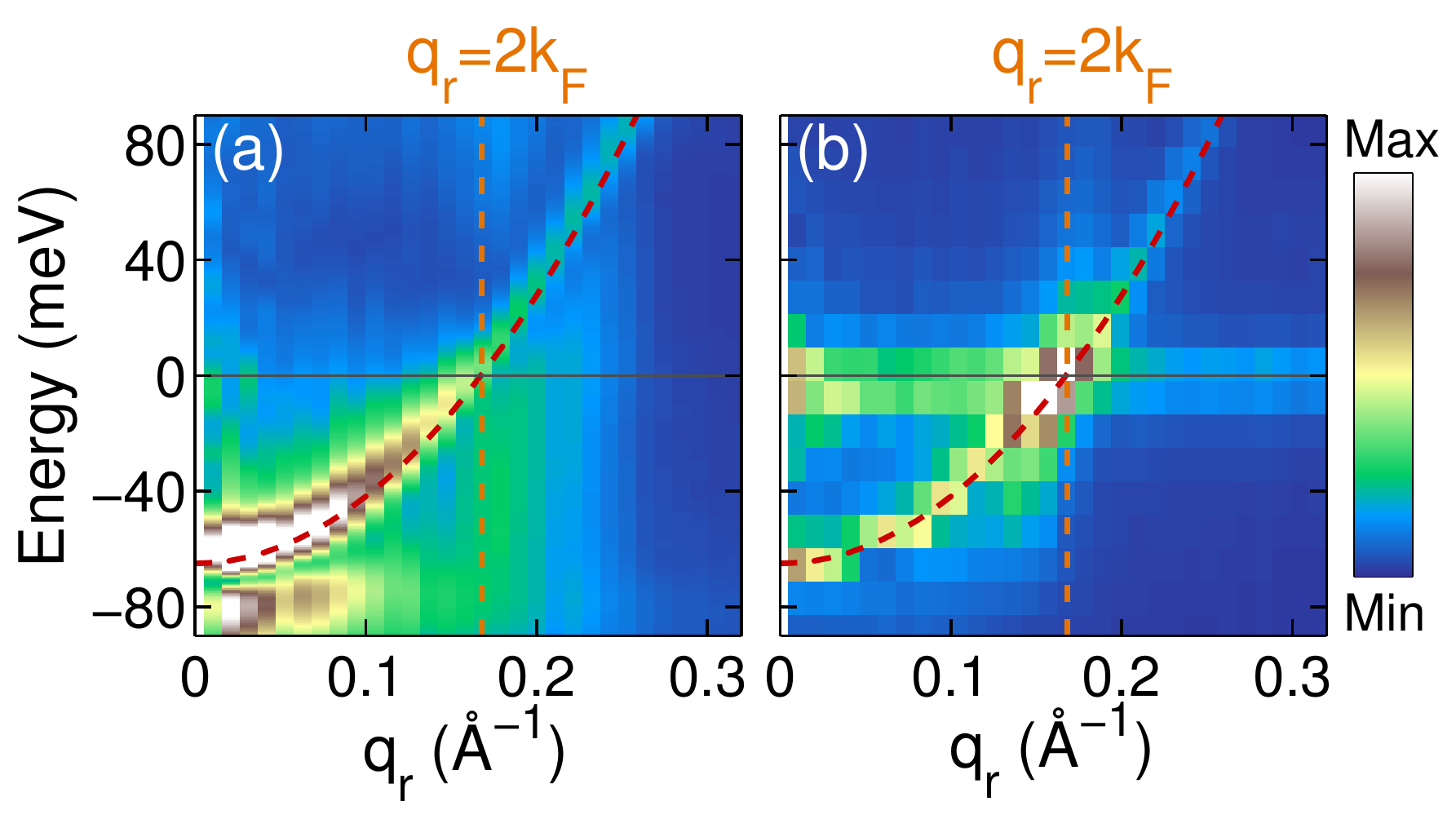}
\caption{Energy dispersion in \textbf{q}-space resulting from two different measurement techniques, spectroscopic grids and constant current lock-in maps. Though the surface state dispersion is clear in both plots the overall intensity differs significantly. Horizontal and vertical lines indicate $E_F$ and $q_F$ respectively and the dashed parabola comes from fitting the effective electronic mass to a free electron model. (a) Spectroscopic grid from a 60 x 60 nm$^2$ area with set point bias of $V_s$ = 100 mV and $I_t$ = 100 pA. (b) Constant current maps taken with a lock-in amplifier over a 60 x 60 nm$^2$ area and a current $I_t = 100$ pA. Lock-in parameters: $f = 1.017$ kHz and $V_{mod} = 5$ mV.} 
\label{Fig2}
\end{center}
\end{figure}

To probe this in more detail, grid measurements were acquired using different bias voltages to stabilize the tip height at each pixel.  We had previously noticed that stabilization biases, $V_s$, corresponding to energies well below the onset of the surface state band, $\mu$, showed no prominent features other than the parabolic dispersion [17], thus minimizing the Òset point effectÓ.  Figure 3 shows each notable case: $eV_s < \epsilon(0)$, $E_F >$ $eV_s$ $>$ $\epsilon(0)$, and $eV_s > $ $E_F$.  As previously observed, there is no prominent feature for a stabilization bias below the onset of the band.  For a stabilization bias between the onset of the band and $E_F$, we see a faint, broad feature below 2\textbf{k}$_F$, and for stabilization biases above $E_F$, we see a faint, broad feature above 2\textbf{k}$_F$.  Therefore, for grid measurements, this additional feature depends on the bias used when the tip height is stabilized, with the feature crossing over 2\textbf{k}$_F$ at the Fermi energy, much like the additional feature observed for dI/dV maps.

\begin{figure}[htbp]
\begin{center}
\includegraphics[width=8.4cm]{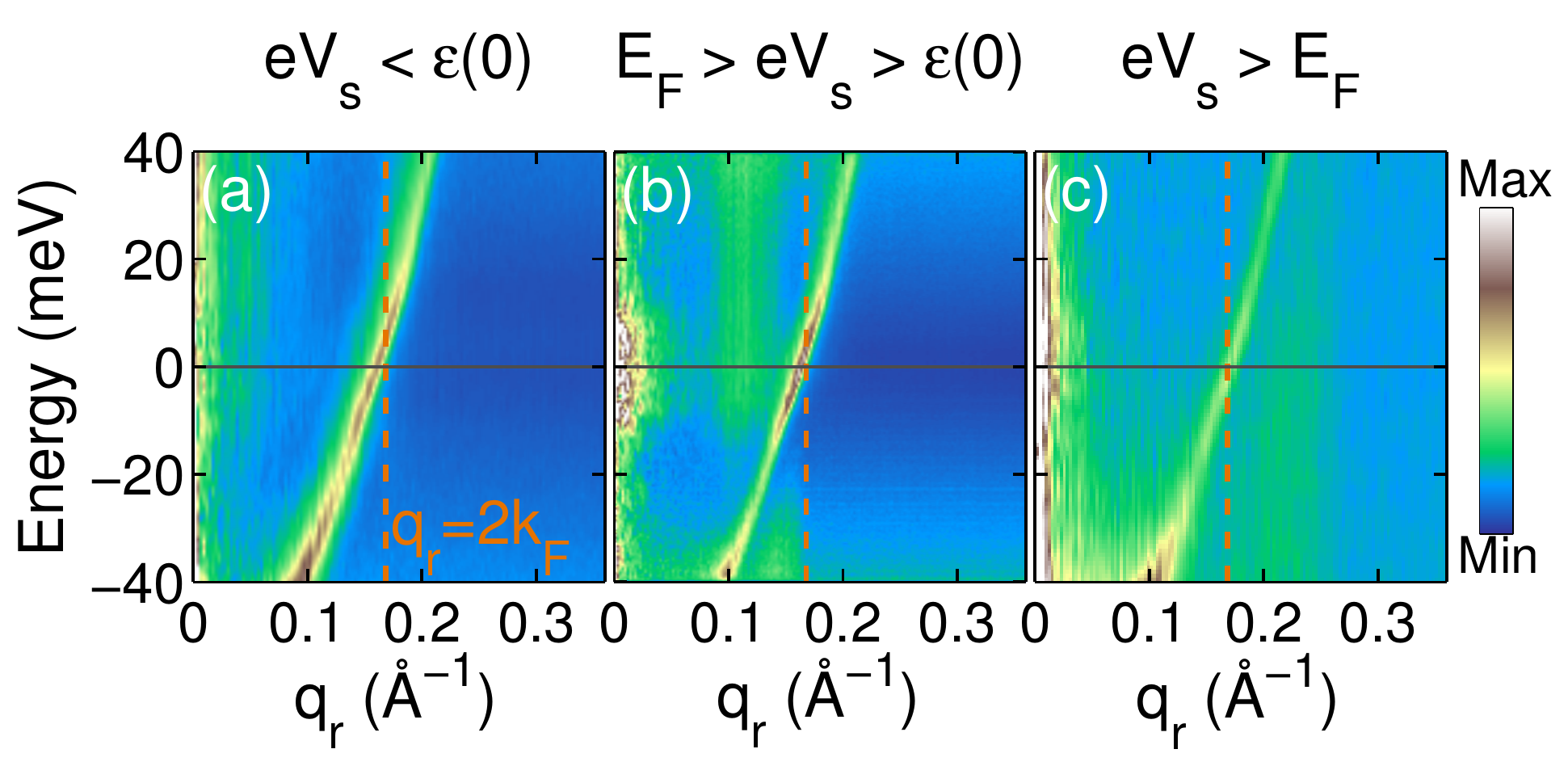}
\caption{Comparison of three spectroscopic grids with three different set point conditions. (a) Grid with $V_s = -100$ mV, $I_t = 100$ pA, real space size 239 x 239 nm$^2$ with 380 x 380 pixels (b) Grid with $V_s = -40$ mV, $I_t$ = 540 pA, real space size 280 x 280 nm$^2$ with 400 x 400 pixels (c) Grid with $V_s = 100$ mV, $I_t = 100$ pA, real space size 240 x 240 nm$^2$ and 350 x 350 pixels.} 
\label{Fig3}
\end{center}
\end{figure}

To make a more direct comparison between the different acquisition modes, Fourier transformed dI/dV maps were generated at the same energy, $E = 50$ meV, by four different acquisition methods (see Figure 4): a grid with $V_s$ corresponding to the same energy as the energy examined, $eV_s = E = 50$ meV (Figure 4b), a grid with opposite polarity $V_s$ from the energy examined, $eV_s = -E = -50$ meV (Figure 4c), a constant-current dI/dV map acquired with a lock-in amplifier where V$_s$ always corresponds to the energy examined, $eV_s = E =$ 50 mV (Figure 4d), and a constant-height dI/dV map acquired with a lock-in amplifier where the tip height is stabilized only at the first pixel of the image (Figure 4e). The sharp peak seen in all four measurements at $q_r = 0.22$ \AA$^{-1}$ corresponds to the intraband scattering of the surface state at 50 meV.  The grid with $V_s = 50$ mV is nearly identical to the constant current dI/dV map; this is expected if the additional feature is a set-point effect since the feedback is stabilized at each point with the same parameters for both measurements.  In addition to the ring corresponding to the expected scattering across the surface state band at $q_r = 0.22$ \AA$^{-1}$, there is a second relatively sharp feature between \textbf{q} = 2\textbf{k}$_F$ and the surface state band, indicated by the blue arrow in Figure 4a).  However, for the grid acquired with $V_s= -50$ mV, while the intraband scattering at $q_r = 0.22$ \AA$^{-1}$ remains the same, there is now a much broader feature centered below 2\textbf{k}$_F$ (indicated by the red arrow), and the feature seen with $V_s = +50$ mV (blue arrow) is no longer observed.  A constant height dI/dV measurement was also acquired, where the tip height is stabilized only in one position at the beginning of the measurement. Although the data has a larger low-frequency background, no clear secondary features are observed. Each of the secondary features appear above the background level of the other line cuts indicating that each are in fact additional features tied only to the measurement mode and parameters. The lack of any secondary features in the constant height measurement, along with the dependence of the additional features on the bias used to stabilize the tip height at each position for grids and maps, point to an influence of the spatially varying tip-height on the dI/dV measurement.

\begin{figure}[htbp]
\begin{center}
\includegraphics[width=8.cm]{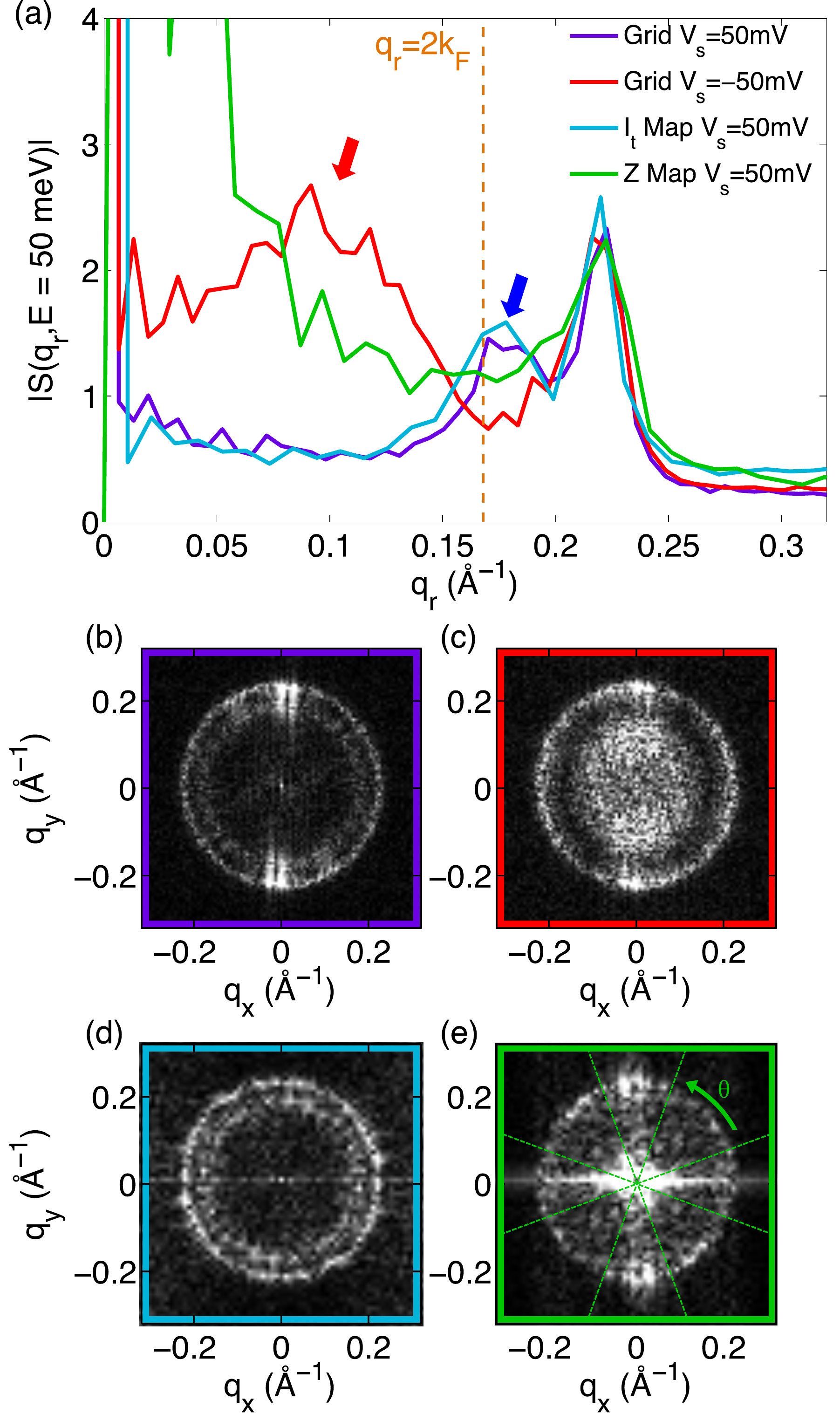}
\caption{The effect of stabilization bias on the observed QPI pattern. (a) Line cuts in \textbf{q}-space taking an angular average and comparing between two grids with different stabilization bias', a constant current map, and a constant height map. (b-e) FT-STS at $E = 50$ meV from (b) a spectroscopic grid with $V_s = 50$ mV, $I_s = 100$ pA, (c) a spectroscopic grid with $V_s = -50$ mV,  $I_s = 100$ pA, (d) constant current map at $V_s = 50$ mV, $I_s= 100$ pA, and (e) constant height map at $V_s = 50$ mV, and initial current $I_s = 100$ pA. For the constant height data a restricted azimuthal average similar to that described in the methods for grid measurements was used to reduce the influence of a step edge running across the top of the image and some additional artifacts introduced by applying a line-by-line subtraction to account for z-drift.} 
\label{Fig4}
\end{center}
\end{figure}

When the tip is stabilized at each pixel, a constant current condition is met by the feedback circuit.  That constant current condition depends on the integrated density of states, convolved with the transmission function of the tunnel junction.  
Since the density of states varies with both position and energy, the constant current topography will contain spatial modulations due to the electronic structure that depend on the bias applied, modulating the physical tip-sample separation.  As the dI/dV signal also contains the transmission function, which depends on tip-sample separation, it is perhaps not surprising that extraneous features are observed in FT-STS that depend on the energy used to stabilize the tip height.  This dependence of the dI/dV on the tip height now explains the differences between ÒgridÓ and constant current dI/dV ÒmapÓ measurements: for a ÒgridÓ only one stabilization bias is used for all energies probed, so a non-dispersing feature either above (positive $V_s$) or below (negative $V_s$) 2\textbf{k}$_F$ is observed, but for constant current dI/dV ÒmapsÓ, the stabilization bias follows the energy being probed generating a secondary feature that disperses, crossing 2\textbf{k}$_F$ at $E_F$ (see Figure 2b).

To understand this effect, we now examine the influence of the spatially varying tip height on the dI/dV signal.  Whenever the feedback circuit is engaged, the STM tip is stabilized to a particular height above the surface, $z_s$, determined by a user specified bias, $V_s$, and tunnelling current set point, $I_s$. In the zero temperature, one-dimensional limit this tunnelling current, $I_s = I(x,y,z_s,V_s)$,  can be approximated by \cite{Hellenthal2013}

\begin{eqnarray}
I_s = \int_0^{eV_s} \rho(x,y,E) \rho_{t}(E-eV_s) T(z_s,V_s,E) dE,
\end{eqnarray}

\noindent where  $\rho$ is the sample density of states, $\rho_t$ is the tip density of states, and $T$ is the tunnelling barrier transmission coefficient. Following previous work \cite{Hellenthal2013,Ziegler2009} we assume a trapezoidal tunnelling barrier and estimate $T(z_s,V_s,E)$ using the WKB approximation which gives

\begin{eqnarray}
T(z_s,V_s,E) = \exp \Bigg( -z_s \frac{2\sqrt{2m}}{\hbar} \sqrt{\phi + \frac{eV_s}{2} - E} \Bigg),
\end{eqnarray}

\noindent where $\phi$ is the effective height of the tunnelling barrier and $m$ is the free electron mass. We take $\phi = 4.65$ eV, the average of the work function of the W tip ($\phi = 4.55$ eV) and the Ag sample ($\phi = 4.74$ eV) \cite{Li1997}. In the low bias approximation we follow Koslowski \emph{et al}. \cite{Koslowski2007} and let $T(z_s,V_s,E) \approx T(z_s,V_s)$. Inserting this into the equation for $I_s$

\begin{eqnarray}
I_s = \int_0^{eV_s} \rho(x,y,E) \rho_{t}(E-eV_s) T(z_s,V_s) dE \\
I_s =  e^{\big( -z_s \frac{2\sqrt{2m \phi}}{\hbar} \big)} \int_0^{eV_s} \rho(x,y,E) \rho_{t}(E-eV_s) dE.
\end{eqnarray}

\noindent Rearranging to solve for $z_s$

\begin{eqnarray}
z_s = - \frac{\hbar}{2\sqrt{2m \phi}} \ln \Bigg( \frac{I_s}{\int_0^{eV_s} \rho(x,y,E) \rho_t(E-eV_s) dE} \Bigg).
\end{eqnarray}

This equation holds the key to understanding how the set point parameters influence the FT-STS results differently for spectroscopic grids, constant current maps, and constant height maps. In grid acquisition $z_s = z(x,y,V_s,I_s)$ and is set by the feedback at each pixel based on the values of V$_s$ and I$_s$. This makes the tip sensitive to lateral variations in the LDOS but since V$_s$ remains the same at every pixel this at most introduces a single spatial frequency corresponding to a non-dispersing feature in \textbf{q}-space. This feature appears at approximately the average of all scattering \textbf{q} values between 0 and eV$_s$ as it is related to the integrated LDOS. This is not the case for constant current maps where the stabilization bias is tied to the map energy E for each map. This means that $z_s = z(x,y,V_s = E/e, I_s)$ where $V_s$ varies, changing the spatial features convolved into the dI/dV measurement. Since $z_s$ contains periodic spatial modulations at approximately the average of all q values between 0 and V$_s$ = E/e, this leads to the presence of dispersing features in the FT-STS pattern. Constant height maps, on the other hand, have no sensitivity to lateral variations of the LDOS. In a constant height map the tip height is set at one position, x$_s$, y$_s$, at the start of the map and then the feedback is disengaged, excluding the possibility of spatially dependent, feedback induced artifacts ie. $z_s = z(x_s,y_s,V_s = E/e, I_s)$.

To properly simulate the contribution to the FT-STS from set point effects we analyze the derivative of the simulated $I(x,y,z,V)$ with respect to the applied bias, to obtain the dI/dV. Taking the full derivative of $I(x,y,z,V)$ gives two terms with a $z_s$ dependence \cite{Hellenthal2013}

\begin{eqnarray}
\frac{dI(x,y,z_s)}{dV} \propto e \rho (eV) \rho_{t} (0) T(z_s) - \frac{\sqrt{2me}}{2 \hbar \sqrt{\phi}} z_s I(z_s,V).
\end{eqnarray}

\begin{figure}[htbp]
\begin{center}
\includegraphics[width=8.1cm]{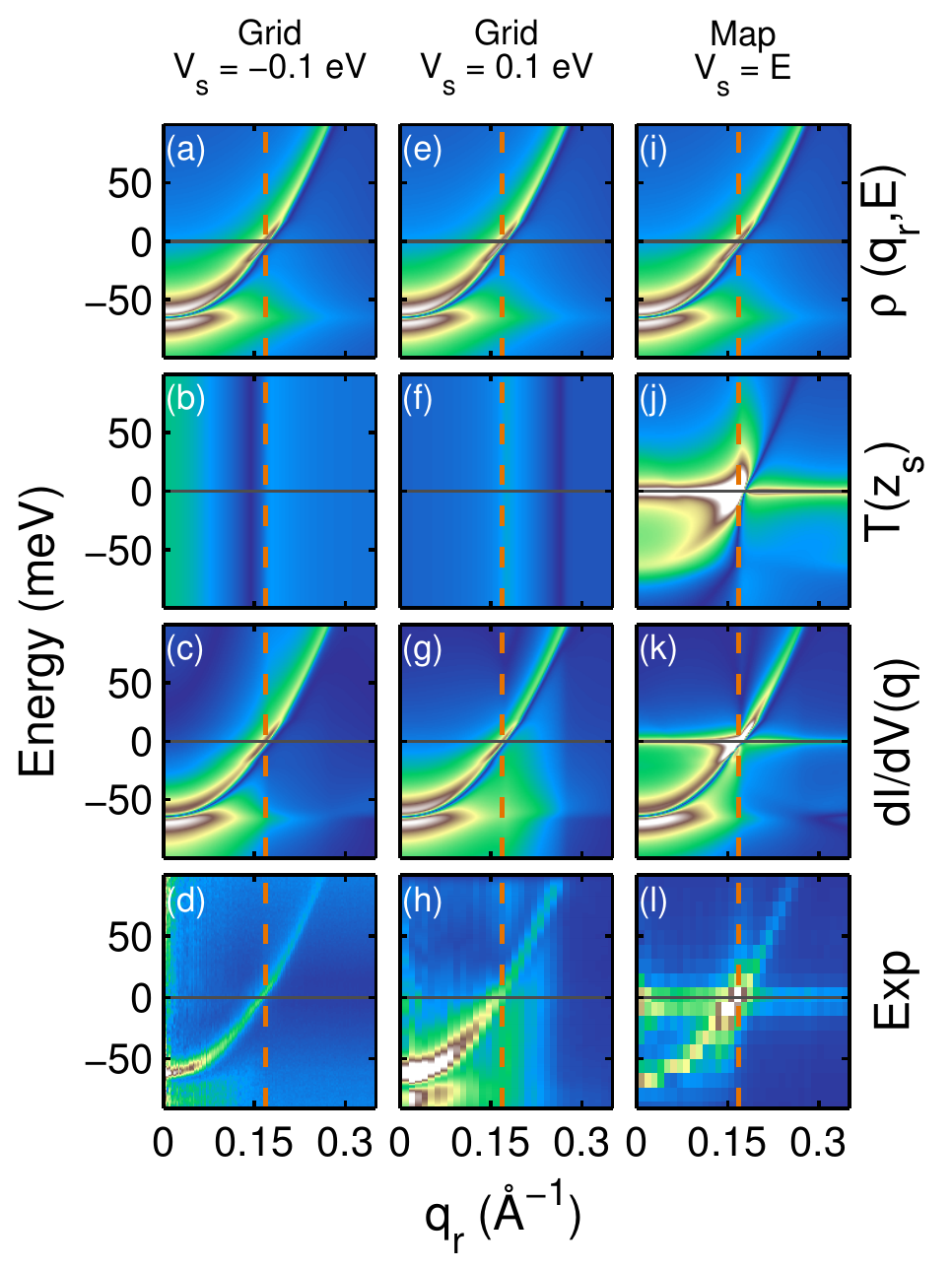}
\caption{Comparison of T-matrix simulations applying the analytical expressions derived here with the experiments for different stabilization bias conditions. Scaled colour intensity is the same across each row. The leftmost column with a)-c) simulates a grid measurement with V$_s = -100$ meV and compares it to an experimental dispersion in d) with the same V$_s$. The middle column e)-g) simulates a grid measurement with V$_s = 100$ meV with the experimental equivalent shown in h). The right column i)-k) shows constant current acquisition mode where the set point energy matches the scan energy as in constant current maps in l). The first row shows the LDOS in q-space as determined from T-matrix simulations. The second row shows the calculated $T(z_s)$ of this LDOS with b) V$_s = -100$ meV bias, f) V$_s=100$ meV bias, and j) V$_s$ varying with the energy probed. The third row shows the product of the LDOS with $T(z_s)$, which qualitatively agrees with the experimental data shown in the fourth row. Note: Energy and \textbf{q} resolution varies between the measurements.} 
\label{Fig5}
\end{center}
\end{figure}

\noindent Comparing the magnitude of these terms using the work function and free electron mass the first term dominates under the experimental conditions considered here, consistent with Li \emph{et al}. \cite{Li1997} 

Using these analytic results, the effect on the FT-STS pattern of the different measurement modes and their set point artifacts can be simulated, expanding on previous work describing the effect of real-space oscillations of $T(z_s)$ in one-dimenstion \cite{Li1997,Hellenthal2013,Koslowski2007,Hormandinger1994,Ukraintsev1996,Pronschinske2011} and on molecules in two-dimensions. \cite{Ziegler2009}

As a basis for the expected FT-STS dispersion, T-matrix simulations described in detail elsewhere \cite{Grothe2013} were used to model scattering of the Ag(111) surface state.  These scattering simulations provided a theoretical density of states in \textbf{q}-space, $\rho$(\textbf{q},E), which was Fourier transformed into a real space to give $\rho$(\textbf{r},E) and used to calculate $T(z_s)$, $I(x,y,z,V)$, and $dI/dV$ under different set point conditions from the analytic expressions derived above. $T(z_s)$ and $dI/dV$ were then Fourier transformed back into \textbf{q}-space and compared with the original density of states, $\rho$(\textbf{q},E). Figure 5 contrasts each case and compares the simulation to the experimental data. The agreement between theory and experiment is qualitatively very good and shows that the acquisition mode dependent features are related to variations in $T(z_s)$. The strongest artifacts are introduced for the constant current maps, while the best match with the underlying LDOS is acquired by a grid with a set point bias below the onset of the surface state.

\section{Discussion \& Conclusions}

As can be seen from the simulations, the dominant factor yielding different results for different measurement conditions arises from the transmission function $T(z_s)$ term.  In the combined dI/dV simulation of the FT-STS dispersion one can see the main features previously described above: for grids with a V$_s$ above $E_F$ there is a clear vertical, non-dispersing line above 2\textbf{k}$_F$, for grids with a V$_s$ below the onset of the band there is only a weak very broad non-dispersing feature in $T(z_s)$ with little influence on the $S(q_r,E)$, and for the series of maps with constant current feedback at each energy there is a dispersing feature which crosses 2\textbf{k}$_F$ at $E_F$ as well as an overall increase in intensity near $E_F$ arising from strong variations in $T(z_s)$. The positions of the features observed in a grid measurement with $V_s = 100 $ meV and a series of constant current and constant height maps are shown in Figure 6 to summarize the potential artifacts. As can be seen, all three consistently reproduce the parabolic intraband scattering dispersion. The grid produces a vertical artifact feature, but only the constant current maps generate a dispersing artifact feature. 

\begin{figure}[htbp]
\begin{center}
\includegraphics[width=8.6cm]{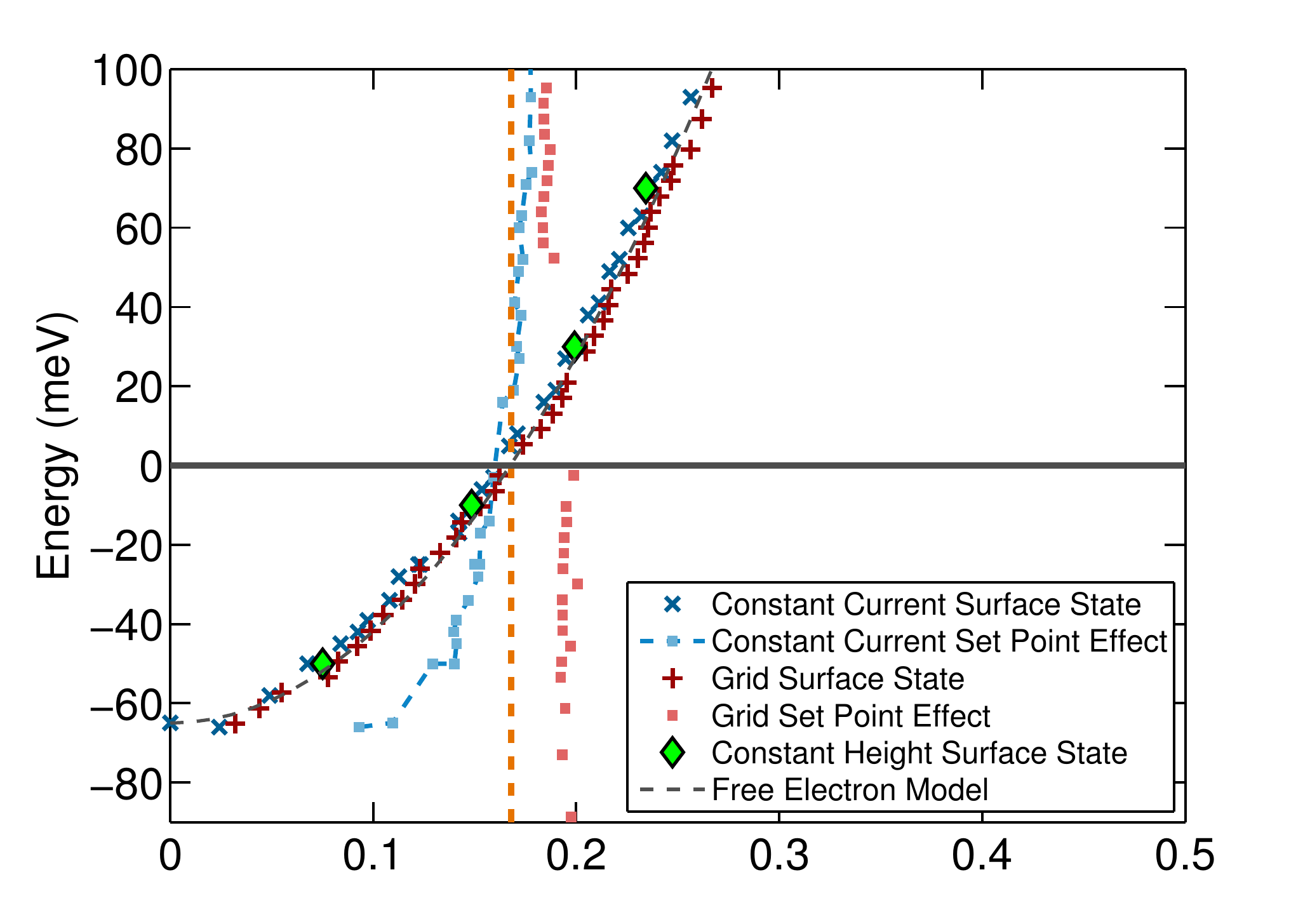}
\caption{Comparison of all features from constant current map, constant height map, and grid $S(q_r,E)$ peak positions illustrating the surface state scattering and set point features. For all three measurement modes the surface state peaks were obtained from Lorentzian fits and in each case the data agrees well with a free electron model of surface state scattering. Set point artifacts were fit with Gaussian functions for both constant current maps and grid data, while the constant height data showed no additional features. Fits of the set-point effect feature for the grid become unreliable close to where this feature crosses the surface state feature and have been omitted here. Only the constant current maps show a dispersing set point peak.} 
\label{Fig6}
\end{center}
\end{figure}

We have demonstrated, through a combination of measurements in different acquisition modes, and simulations of the expected FT-STS patterns for these modes, that artifact features in FT-STS derived dispersions can occur that depend on the set-point conditions used to stabilize the tip height.  The simulations show that this arises from spatial modulations in the transmission function due to variations in $z_s$ at each (x,y) pixel that are dependent on the set-point conditions.  This effect is most pronounced, and most concerning, for measurements acquired by taking dI/dV maps with a simultaneous constant current feedback at each energy, as this produces a relatively strong, dispersing feature.  

Dispersing features with similar $q$-dependence have been observed on (111) noble metal surfaces using constant current maps and attributed to a number of different sources. Petersen \emph{et al.} \cite{Petersen1998} first reported a secondary scattering ring and ascribed this to scattering across a neck in the bulk Fermi surface in measurements of Au(111) and Cu(111). Schouteden \emph{et al.} \cite{Schouteden2009}, performed further measurements of the Au(111) surface state and attributed the lack of dispersion of the bulk band above $E_F$ to inelastic electron relaxation rates. Most recently, Sessi \emph{et al.}  \cite{Sessi2015} demonstrated that the secondary dispersion is not compatible with the position of the bulk bands in Au(111), Cu(111), or Ag(111) and instead attributed the secondary features to an acoustic surface plasmon dispersion. Our data shows that a feature following this same dispersion arises from an artifact of the constant current measurement mode. We note that in constant height maps this set point effect due to modulation of the tunnel barrier is not present. However, for sufficiently high tunnelling current Equation 7 also predicts that a secondary dispersing feature should appear arising from the second term, even in constant height map data. Although we did not observe features in constant height maps related to this on Ag(111), this is a possible explanation for the secondary dispersing features observed in constant height maps on Cu(111) by Sessi \emph{et al.} \cite{Sessi2015}. If so, it should be linearly dependent on the tunnelling current, thus providing a way to test whether a secondary dispersing feature in constant height maps is caused by the physics of the tunnel junction or many-body effects in the sample.

Lastly, we note that for grid measurements a choice of stabilization bias below the onset of the band showed a very weak influence with no distinct features, providing a way to avoid these effects without resorting to demanding constant height measurements, or a more elaborate program of returning the tip to the same location to reset the height for each measurement pixel, as has been done for AFM measurements \cite{Mohn2011}.  These results urge caution in the field; features in QPI require careful consideration, and artifacts can arise depending on the measurement mode that may obscure or masquerade as physical processes in the sample.

\ack

The authors thank D. Bonn for valuable discussions. This work was supported by the NSERC Discovery Grant Program (No. 402072-2012), NSERC CREATE Program (No. 414110-2012), UBC, the Canada Research Chairs Program (S. A. Burke), and the NSERC PGS program (A. J. Macdonald). 

\section*{References}

\bibliographystyle{iopart-num}
\bibliography{references}

\end{document}